\documentclass[conference]{IEEEtran}

\ifCLASSOPTIONcompsoc
  \usepackage[nocompress]{cite}
\else
  \usepackage{cite}
\fi

\usepackage{graphicx}

\usepackage{color}
\definecolor{mygrey}{gray}{0.50}

\begin{document}
\title{Classification of Hybrid Quantum-Classical Computing}


%

\author{\IEEEauthorblockN{Frank Phillipson}
\IEEEauthorblockA{TNO\\
Dpt. Applied Cryptography \& Quantum Algorithms \\
The Hague, The Netherlands\\
\\
Maastricht University\\
School of Business and Economics\\
Maastricht, The Netherlands}
\and
\IEEEauthorblockN{Niels Neumann and Robert Wezeman}
\IEEEauthorblockA{TNO \\
Dpt. Applied Cryptography \& Quantum Algorithms\\
The Hague, The Netherlands}
}


%

\maketitle

\begin{abstract}
As quantum computers mature, the applicability in practice becomes more important. Many uses of quantum computers will be hybrid, with classical computers still playing an important role in operating and using the quantum computer. The term hybrid is however diffuse and multi-interpretable. In this work we define two classes of hybrid quantum-classical computing: vertical and horizontal. The first is application-agnostic and concerns using quantum computers. The second is application-specific and concerns running an algorithm. For both, we give a further subdivision in different types of hybrid quantum-classical computing and we coin terms for them. 
\end{abstract}

\begin{IEEEkeywords}
hybrid quantum computing, classification, hybrid quantum algorithm, workflow
\end{IEEEkeywords}

\IEEEpeerreviewmaketitle

\section{Introduction}
Quantum computing is the technique of using quantum mechanical phenomena such as superposition, entanglement and interference for doing computational operations. The type of devices which are capable of doing such quantum operations are still being actively developed and named quantum computers. We distinguish between two paradigms of quantum computing devices: gate-based quantum computers and quantum annealers. A gate-based quantum computing system uses basic quantum circuit operations on qubits, similar to the classical operations on regular bits, that can be put together in any sequence to form algorithms. A quantum annealer brings a collection of qubits into an equal superposition and then applies a problem specific magnetic field. The qubits will interact under this magnetic field and move towards the state with the lowest energy, which encodes the solution of an optimisation problem.

Theory predicts that quantum computers will solve specific problems much faster than classical computers. Where classical computers have been under development for decades and are therefore quite mature, quantum computers are still in the early stages of development. They are not yet capable of solving real world problems, due to the low number of qubits and their unstable nature causing noise, errors and loss of information. This current state of quantum computers is called the noisy intermediate-scale quantum (NISQ) era introduced by Preskill~\cite{preskill2018quantum}, where quantum computers have 50-200 qubits and their noise places serious limitations on their capabilities. 
Only recently IBM passed the 100-qubit barrier for the first time with its $127$-qubit Eagle processor and they plan a $1{,}000$-qubit chip, the Condor, in 2023~\cite{IBM_Qroadmap:2020}.

Other aspects than number of qubits affect the capabilities of quantum computers in practice, such as parallelism of operations and the topology layout of the qubits. Researchers are investigating innovative ways to solve valuable problems using already available NISQ systems and to achieve quantum advantage by demonstrating a significant performance advantage over today’s classical computers. The latter was only quite recently claimed for the first time for an artificially created problem~\cite{arute2019quantum}. As quantum computers mature in the coming years, their computational power will increase and they can be applied in more settings and actually provide help in specific practical areas. 

\begin{figure*}[ht]
    \centering
    \includegraphics[width=\textwidth]{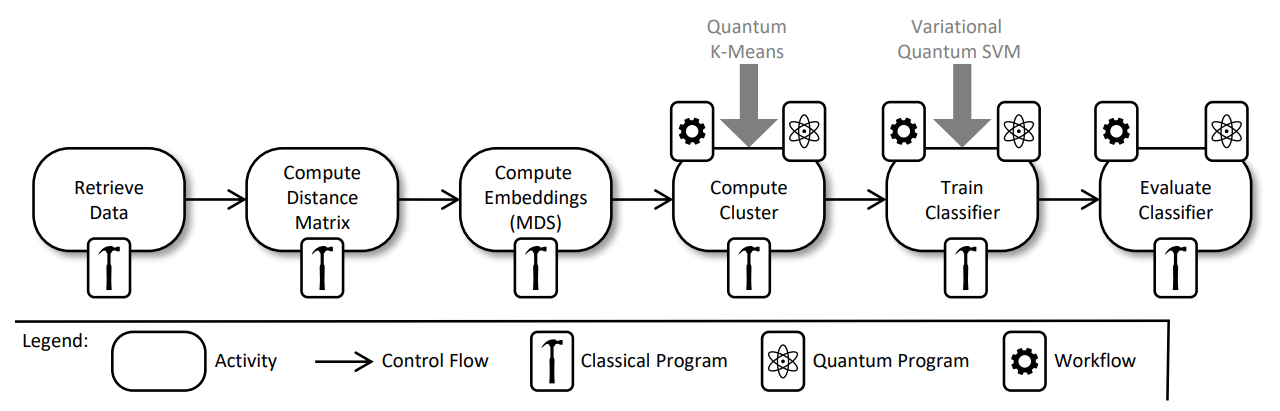}
    \caption{An example workflow of a hybrid quantum-classical application~\cite{weder2021hybrid}. Quantum computers perform only a small part of the computations.}
    \label{fig:wf1}
\end{figure*}

For the first practical applications in the near future, quantum computers will only execute a small part of a larger total workflow where a classical computer executes the other steps. See for example the workflow of retrieving data, training a classification algorithm and evaluating the obtained classifier~\cite{weder2021hybrid}, as shown in Fig.~\ref{fig:wf1}. In this example only a small part of the workflow is performed on a quantum computer. Already in 2005 such a combination was described in literature and was named hybrid quantum computing that ``combine both classical and quantum computing architectures in order to leverage the benefits of both"~\cite{lanzagorta2005hybrid}. 

Murray Thom, the Vice President of Software and Cloud Services at D-Wave, compared this with jets and "normal" vehicles: "Consider that while jet airplanes transformed the way we travel long distances, we still need vehicles that take us to our front door." Thus, "quantum applications will always and only be hybrid"\cite{Hybrid}.

This term, hybrid quantum-classical computing, was and is used in various contexts, each time used for (slightly) different settings. This is confusing for many researchers and practitioners in quantum computing, both on the application side as on the hardware side. The interest in quantum computing causes an enormous amount of research output\footnote{Google scholar already gives 2090 results for the search on 'hybrid quantum-classical computing' for the period January-October 2022.} and, resulting, survey papers. One example of this kind is \cite{de2022survey}, where a survey of NISQ era hybrid quantum-classical machine learning research on hybrid quantum-classical systems is given, without explaining what is meant by hybrid quantum-classical exactly. This is totally left to the reader. Our goal is to provide a classification framework where authors can refer to, to make the scope of their contribution more clear to the reader. For this, we connect quantum research with classical computer science and workflow terminology. So, many of the proposed terminology in this paper exists already in computer science, think of compilation between computer languages, and workflow research, think of decomposition and activities. Our contribution is bringing them together in a clear framework for hybrid quantum-classical computing.

In this contribution we aim to describe clearly the various contexts hybrid quantum-classical computing can have in literature and to name these different approaches clearly and appropriately. In Section~\ref{sec:literature} we give an overview of various forms of hybrid quantum-classical computing that can be found in literature. Next, in Section~\ref{sec:types_hybridQC}, we distinguish a number of different types of hybrid quantum-classical computing from this overview and provide examples for each type. We end this paper with some conclusions and ideas for further research. 

\section{Literature\label{sec:literature}}
We consider a global situation where we have a collection of computational tasks in which both the quantum computer and the classical computer are used. As such, hybrid forms of computing that allow for both discrete and continuous variables~\cite{lloyd2003hybrid} and hybrid quantum-classical models of molecules in chemical and biological studies~\cite{henelius2008hybrid} are out of scope. We do not try to give an  exhaustive overview of all research done on this topic. Our goal here is to give an overview, based on some examples, of the various meanings of the term hybrid quantum-classical computing in literature. This overview will be the basis of the proposed classification later on in this work. 

Lanzagorta and Uhlmann presented one of the first hybrid algorithms that used both classical and quantum computers~\cite{lanzagorta2005hybrid}. Later, research appeared on computing schemes and architectures to optimise the interactions between the different type of computers when executing hybrid quantum algorithms. A first example presents a candidate framework to analyse hybrid computations by fully integrating the quantum and classical resources and processes used for measurement-based quantum computing, where the feed-forward of classical measurement results is an integral part of the quantum design~\cite{Jozsa_MeasurementBasedQC:2005,horsman2009hybrid}. A second example proposes a quantum co-processor to accelerate a specific subroutine of a larger task. This is most often seen as the main reason for hybrid algorithms, for example in~\cite{possignolo2012quantum,bravyi2016trading,li2017hybrid,mccaskey2020xacc}. The work by Li et al.~\cite{li2017hybrid} results in a system-level software infrastructure for hybrid quantum–classical computing. Endo~\cite{endo2019hybrid,Endo:2021} indicates that for early quantum applications, a large portion of the computational burden is performed on a classical computer and hence fully coherent deep quantum circuits are not required. As the quantum computer takes on more computational load, noise of the quantum computer will result in more errors, which will have more impact on the total calculation. This in turn requires qubits of higher quality and error mitigation routines to suppress noise.  

An important type of hybrid computing appeared with the introduction of Variational Quantum Algorithms (VQA). VQAs use a classical optimiser to train a parameterised quantum circuit and provide a framework to tackle a wide array of tasks, as shown in the extensive overview by Cerezo et al.~\cite{cerezo2021variational}. Examples include finding ground states and excited states of molecules, optimisation, solving linear systems of equations and machine learning. The first VQA, the variational quantum eigensolver (VQE) algorithm of~\cite{peruzzo2014variational}, appeared in 2014. Some papers see this group of algorithms as ``a novel class of hybrid quantum-classical algorithms", without explicitly diving in other groups of algorithms~\cite{calude2015guest}.
Also, ~\cite{corcoles2019challenges} sees this class of algorithms as a specific example of hybrid quantum-classical computing in a noisy environment.
In~\cite{weder2020quantum} the main reason given for hybrid quantum-classical computing is the size of the problems in combination with the available hardware. They distinguishes two types of hybrid computing: 1) Pre- or post-processing of a quantum computation on a classical computer. Examples are algorithms by Shor~\cite{Shor:1997} and Simon~\cite{Simon:1997} that use classical post-processing. 2) Algorithms that perform multiple iterations of quantum and classical computations. Thereby, the output of the quantum computation is improved in each iteration until the result reaches the required accuracy. An example they give using this approach is the quantum approximate optimisation algorithm (QAOA)~\cite{FahriGoldstoneGutmann:2014}.
\begin{figure*}[ht]
    \centering
    \includegraphics[width=\textwidth]{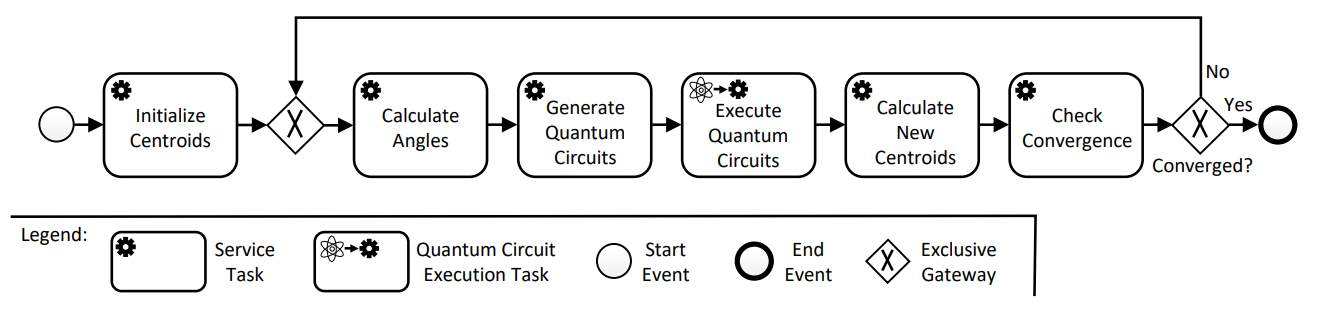}
    \caption{The sub-workflow of the ``Compute Cluster activity" shown in Fig.~\ref{fig:wf1}~\cite{weder2021hybrid}.}
    \label{fig:wf2}
\end{figure*}

In the works~\cite{calude2015guest,calude2018quassical}, the term ‘quassical’ computing is coined and motivated by ``Classical computing and quantum computing have obvious complementary strengths, so instead of opposing them it might be better to combine them into a new type of computing." They give two reasons for the combination: First, most quantum computing algorithms ``require some preliminary classical pre-processing to shape the problem into one the quantum computer can recognise and then to receive the data returned by the [quantum computer] and shape it into the answer the engineer needs." Second, ``all the quantum computers we have heard of are designed as cyber-physical systems, quantum mechanical systems controlled by digital controllers", meaning they are quassical in a trivial sense. In this light, you can also think of the classical steps needed to transform a quantum circuit to an execution as performed for example by an openQL framework \cite{khammassi2021openql}. They expect that this combination will stay, also when the quantum computer is in full maturity.   The first remark is also mentioned by~\cite{edwards2020towards}, who indicate that ``while hybrid algorithms and platforms may just be the best first step, it is reasonable to assume that quantum applications will always be hybrid", for example, by the need of a pre-processing step which prepares data for a quantum algorithm or a post-processing step which handles data coming from a quantum algorithm.

Another view on hybrid algorithms is given by the idea that a problem or circuit that is too big to be executed on (noisy) quantum processors of intermediate size, is partitioned automatically into smaller parts that are evaluated separately. Suchara et al.~\cite{suchara2018hybrid} suggest such an approach for gate-based quantum computers: ``We advocate using a hybrid quantum-classical architecture where larger quantum circuits are broken into smaller sub-circuits that are evaluated separately, either using a quantum processor or a quantum simulator running on a classical supercomputer." 
The D-Wave hybrid solvers should also be seen in this light. They offer the functionality to partition the problem into smaller pieces that fit the current chip size and are solved sequentially. The outcomes of the subroutines form the resulting (probably sub-optimal) solution. They also provide hybrid approaches, where multiple branches of a process solve the problem, some of them with a classical solver, others using the QPU, and then return the first or best solution~\cite{booth2017partitioning}. Here, the quantum and classical computers compete in parallel to find a solution to the problem. In~\cite{chiscop2020hybrid} the D-Wave Kerberos solution is used, a hybrid built-in sampler, that combines Tabu search, simulated annealing, and D-Wave sub-problem sampling on problem variables that have high-energy impact.

\section{Types of hybrid computing\label{sec:types_hybridQC}}
To make a clear distinction between all the different views on hybrid quantum-classical computing, we use the workflow approach presented in~\cite{weder2021hybrid}. They propose workflows to specify the (partial) order of a collection of activities needed to execute a hybrid quantum-classical application and combine this with topologies to reveal the overall structure of hybrid quantum applications. An activity in such a workflow can be further expanded in sub-workflows. Typically, the activities are represented as nodes in a directed graph with the control flow dependencies the directed edges of the graph. Figure~\ref{fig:wf1} shows an example, where the workflow of a hybrid quantum-classical application in quantum machine learning is shown. The presented quantum application performs clustering on a set of input data, and based on the clustering results, trains a classifier for the classification of future data. The ``gear" icon indicates sub-workflows. An example of the ``Compute Cluster" sub-workflow is shown in Fig.~\ref{fig:wf2}. They consider two dimensions of hybrid computing, the vertical provisioning engine and the horizontal workflow-engine, from a software architecture point of view. This insight and the workflow technology are the basis of our proposed classification framework. However, we elaborate further on this and expand it into sub-categories. 

We distinguish two main categories of hybrid computing:
\begin{enumerate}
    \item Vertical hybrid quantum computing: All controlling activities required to control and operate a quantum circuit on a quantum computer, as was the case in classical computing providing compilation and controlling in the stack. An example of a quantum stack can be found in \cite{riesebos2019quantum}. These steps are application-agnostic. 
    \item Horizontal hybrid quantum computing: All operational activities required to use a quantum computer and a classical computer to perform an algorithm. These steps are application-dependent. Here we use the classical workflow terminology as proposed in \cite{ellis1999workflow}.
\end{enumerate}
\noindent We can subdivide both categories further. Note that some use-cases might show signs of more than one type of hybrid quantum-classical computing. It is also important to stress that these main categories are not mutually exclusive. The vertical category is mainly about computing and computing stack, the horizontal category is mainly about algorithms.


\subsection{Vertical hybrid quantum computing}
\begin{figure}[ht]
    \centering
    \includegraphics[width=\columnwidth]{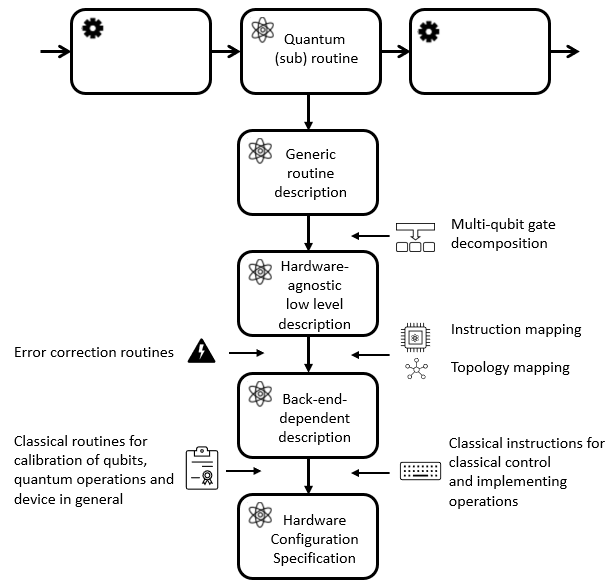}
    \caption{Schematic view on relation between horizontal and vertical dependencies.}
    \label{fig:vert}
\end{figure}

The types of vertical hybrid quantum computing contain the classical steps that have to be taken to let the quantum computer run the quantum routine. A general overview, starting from a single activity in a sub-workflow is depicted in Fig.~\ref{fig:vert}. The specific steps are described in the next sub-sections. These steps are in some way similar to the layers in the OpenQL framework~\cite{khammassi2021openql}, however, we distinguish more between steps that are topology and technology-agnostic and steps that are not.

\subsubsection{Decomposition hybrid}
\begin{figure}[h]
    \centering
    \includegraphics[width=6cm]{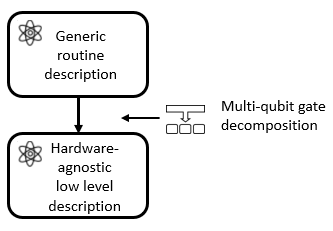}
    \caption{Schematic view on \emph{decomposition hybrid}. A high-level quantum routine is broken down into low-level hardware-agnostic quantum instructions.}
    \label{fig:decomp}
\end{figure}

The workflows of decomposition hybrid consider a higher level algorithm description, which is then decomposed in classical instructions and quantum instructions, see Fig.~\ref{fig:decomp}. A high-level quantum routine is broken down into low-level hardware-agnostic quantum instructions. The higher level algorithm description can be any commonly used classical language, enhanced with classical routines, or a dedicated quantum routine. For gate-based devices, this hardware-agnostic instruction set can for instance include single qubit rotations and some two-qubit gates such as the CNOT-gate and controlled-phase-gates. Libraries can help decomposing quantum instructions to low level hardware-agnostic instructions~\cite{vandenbrink2019vision}.

Examples of decomposition hybrid include decomposing algorithmic instruction to classical instructions and quantum instructions and mapping both to low-level instructions. This includes decomposing high-level instructions not suited for the hardware to lower-level instructions with a more direct mapping to hardware. Classical routines can help with this decomposition~\cite{Kitaev:1997}. The vertical hybrid can be separated into the following parts, which are depicted in Fig.~\ref{fig:decomp}.

\subsubsection{Implementation hybrid}
\begin{figure}[h]
    \centering
    \includegraphics[width=\columnwidth]{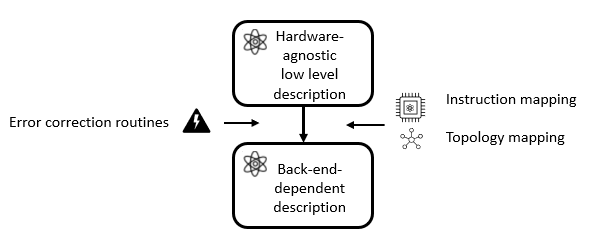}
    \caption{Schematic view on \emph{implementation hybrid}. Low-level quantum instructions are mapped to a specific hardware-backend. This includes creating a time-schedule, assigning operations to qubits and, if necessary, apply error correcting routines.}
    \label{fig:implm}
\end{figure}

The workflows of implementation hybrid consider all steps to map and implement the operations on a quantum computer. This workflow specifically aims to map the hardware-agnostic low-level instructions to hardware-specific instructions, see Fig.~\ref{fig:implm}. This includes both the classical and quantum instructions. 

Examples of implementation hybrid include mapping instructions to hardware. This includes assigning operations to qubits and taking into account the topology of the hardware backend. This workflow also outputs a time-schedule when which hardware instructions should be applied to which specific qubits. If necessary, this workflow also adds error correcting routines together with the classical feedback loop of these routines. 

\subsubsection{Controlling hybrid}
\begin{figure}[h]
    \centering
    \includegraphics[width=\columnwidth]{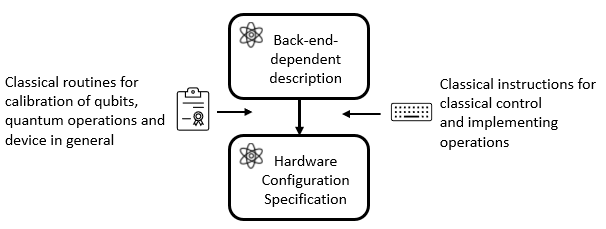}
    \caption{Schematic view on \emph{controlling hybrid}. This considers all steps to operate and control a quantum computer, including the actual mapping of hardware instructions to qubits and calibrating the quantum computer.}
    \label{fig:control}
\end{figure}

The workflows of controlling hybrid consider all steps to operate and control a quantum computer, see Fig.~\ref{fig:control}. Due to the intricate nature of quantum computers, their operations might behave differently over time than intended. Therefore, continuous effort is needed to ensure that quantum computers behave as expected. 

Examples of controlling hybrid include calibration routines of the qubits and of elementary operations on the qubits. 

All of the steps in vertical hybrid quantum computing can include optimisation steps. Sometimes, optimised approximation methods yield better performing implementation then exact full implementations, see, e.g., \cite{schalkers2022learning}.

\subsection{Horizontal hybrid quantum computing}
The types of horizontal hybrid quantum computing distinguish between the variety of orderings quantum and conventional computing steps within an algorithm. 
\subsubsection{Processing hybrid}
\begin{figure}[ht]
    \centering
    \includegraphics[width=7cm]{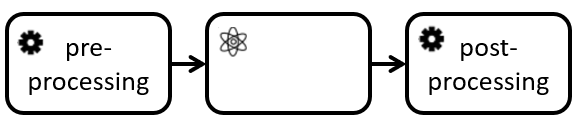}
    \caption{Schematic \emph{processing hybrid} workflow. A single quantum routine with additional classical pre- and post-processing.}
    \label{fig:str_a}
\end{figure}

The workflows of processing hybrid have a single quantum block, combined with classical pre-processing and classical post-processing~\cite{calude2015guest,edwards2020towards}. A schematic representation is shown in Fig.~\ref{fig:str_a}.

Examples of processing hybrid algorithms are the algorithms by Shor and Simon~\cite{suchara2018hybrid}. Another example is the distance based classifier~\cite{Schuld:2017,wezeman2020distance}, where data standardisation and normalisation are the pre-processing steps and translating the measurements to the desired kernel classifier the post-processing step.

\subsubsection{Micro hybrid split}
\begin{figure}[ht]
    \centering
    \includegraphics[width=\linewidth]{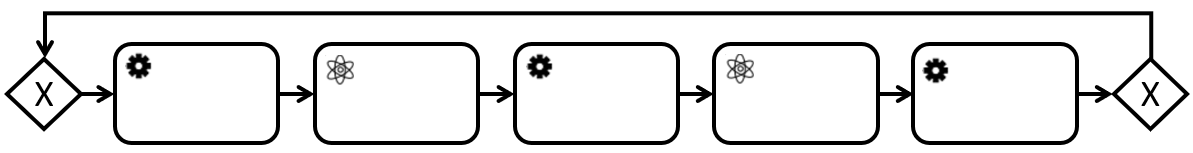}
    \caption{Schematic \emph{micro hybrid split} workflow. A single activity has both classical and quantum operations.}
    \label{fig:str_b}
\end{figure}

The workflows of a micro hybrid split consider a single activity of a larger workflow. The workflow shown in Fig.~\ref{fig:wf2} is a micro hybrid split of Fig.~\ref{fig:wf1}. Within the single activity, some operations are quantum and others are classical, possibly in an iterative fashion. A schematic representation is shown in Fig.~\ref{fig:str_b}. 

Examples of micro hybrid splits are variational algorithms~\cite{calude2015guest}. Measurement-based quantum computing can be seen as a special member of this class, as future measurements and operations depend on previous measurements and classical operations. 

\subsubsection{Macro hybrid split} 
\begin{figure}[ht]
    \centering
    \includegraphics[width=\linewidth]{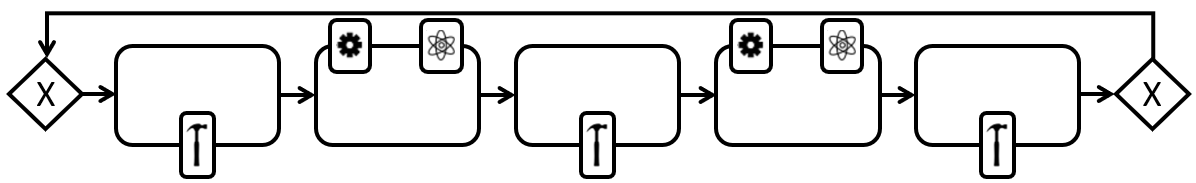}
    \caption{Schematic \emph{macro hybrid split} workflow. Each block is a specific task that can have both quantum and classical operations. A single block can be further subdivided in a micro hybrid split workflow.}
    \label{fig:str_c}
\end{figure}

\begin{figure}[ht]
    \centering
    \includegraphics[width=\linewidth]{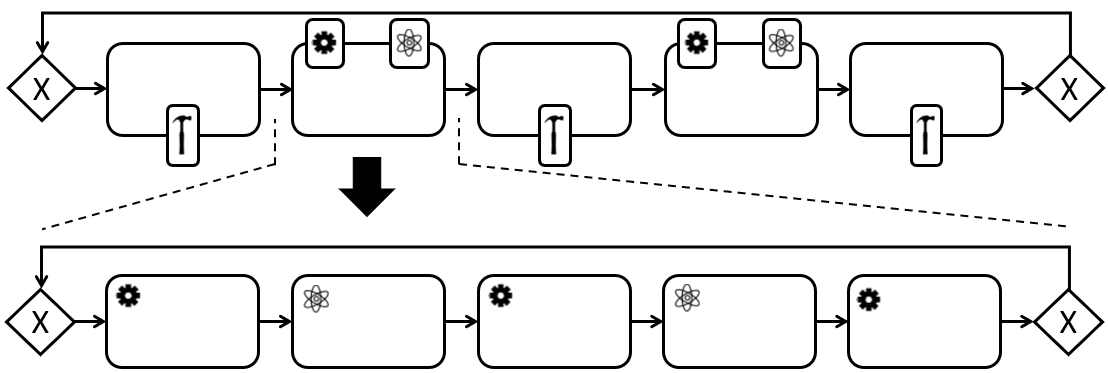}
    \caption{Schematic relationship between \emph{macro hybrid split} and \emph{micro hybrid split} workflow. A block in the \emph{macro hybrid split} can be specified as a \emph{micro hybrid split}.}
    \label{fig:relMM}
\end{figure}

The workflows of a macro hybrid split consider different tasks that belong to different activities within a larger algorithm. The difference with micro hybrid split workflows is small and depends mainly on the granularity with which the workflow is observed: A micro hybrid split in one workflow can be a small part of a larger macro hybrid split workflow. The activities in a macro hybrid split can also iterate and each task can be hybrid in itself. A schematic representation is shown in Fig.~\ref{fig:str_c}. A possible relation between macro and micro hybrid splits is shown in Fig.~\ref{fig:relMM}

Examples of macro hybrid splits are the hybrid quantum machine learning approach in the domain of humanities~\cite{weder2021hybrid} and the workflow shown in Fig.~\ref{fig:wf1}.

\subsubsection{Parallel hybrid} 
\begin{figure}[ht]
    \centering
    \includegraphics[width=6.5cm]{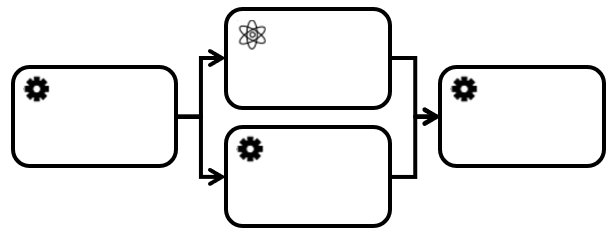}
    \caption{Schematic \emph{parallel hybrid} workflow. A task is processed by multiple independent branches. The answer is returned based on some criteria, for instance, coming from the branch that finishes first.}
    \label{fig:str_d}
\end{figure}

The workflows of parallel hybrid have multiple independent branches to solve a specific problem. Each branch tries to solve the problem independently and the first (or best) solution found is returned. Each branch can use different solvers. A schematic representation is shown in Fig.~\ref{fig:str_d}. 

Examples include the configuration of the D-Wave-hybrid framework, where samples are parsed to four parallel solvers. One branch can for instance be a classical tabu search that either returns with certainty an answer, or is interrupted by another finished branch~\cite{booth2017partitioning,chiscop2020hybrid}. 

\subsubsection{Breakdown hybrid}
\begin{figure}[ht]
    \centering
    \includegraphics[width=\linewidth]{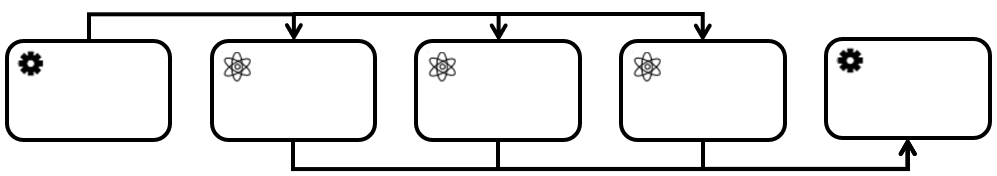}
    \caption{Schematic \emph{breakdown hybrid} workflow. A large problem is decomposed in smaller problems, each of which is run on a quantum computer. The final answer is reconstructed from the partial answers.}
    \label{fig:str_e}
\end{figure}

The workflows of breakdown hybrid consider multiple small parts of a larger problem. The considered problem is too large to solve directly and is hence broken down in multiple smaller parts. Each smaller part is run on a quantum computer sequentially and the final answer is reconstructed from the partial answers. A schematic representation is shown in Fig.~\ref{fig:str_e}. 

Examples are the gate-based approach in~\cite{suchara2018hybrid}, where they advocate using a hybrid quantum-classical architecture where larger quantum circuits are broken into smaller sub-circuits that are evaluated separately and specific options within D-Wave's hybrid solvers~\cite{booth2017partitioning}, where the problem divided into several parts that are solved using classical or quantum annealing approaches. Note that these classes are not disjoint. Here, one of the breakdown parts might be run on a classical computer in parallel with one or more quantum tasks. This would make it a combination of breakdown hybrid and parallel hybrid.

\section{Application}
As indicated in the introduction, the number of papers that are categorised under hybrid quantum-classical computing is enormous. We will not give an exhaustive overview of all papers and their classification. As an example, we selected a few papers from 2022 that have this terms in their title or key words to illustrate the classification.


\begin{table*}[h]
    \centering
    \caption{Example classification of recent papers.}
    \label{tab:class}
    \begin{tabular}{l|l}
Title & Classification \\  \hline
NetQASM-a low-level instruction set architecture for hybrid quantum-classical programs in a quantum internet	\cite{dahlberg2022netqasm}&	Decomposition hybrid\\
QuantumPath: A quantum software development platform \cite{hevia2022quantumpath}&	Decomposition hybrid\\
An LLVM-based C++ compiler toolchain for variational hybrid quantum-classical algorithms and quantum accelerator \cite{khalate2022llvm}&	Implementation hybrid\\
Diversity of hybrid quantum system	\cite{hirayama2022diversity}&	Controlling hybrid\\
Quantify-scheduler: An open-source hybrid compiler for operating quantum computers in the NISQ era	\cite{crielaard2022quantify}&	Controlling hybrid\\
The optimization landscape of hybrid quantum-classical algorithms: from quantum control to NISQ applications	\cite{ge2022optimization}&	Micro hybrid split\\
A hybrid quantum-classical CFD methodology with
benchmark HHL solutions \cite{lapworth2022hybrid}&	Micro hybrid split\\
Considerations for evaluating thermodynamic properties with hybrid quantum-classical computing work flow	\cite{stober2022considerations}&	Micro hybrid split\\
Hyperparameter optimization of hybrid quantum neural networks for car classification	\cite{sagingalieva2022hyperparameter}&	Macro hybrid split\\
Hybrid quantum-classical algorithms for approximate graph coloring	\cite{bravyi2022hybrid}&	Macro hybrid split\\
A hybrid quantum-classical neural network architecture for binary classification	\cite{arthur2022hybrid}&	Macro hybrid split\\
Hybrid quantum-classical algorithm for computing imaginary-time correlation functions	\cite{sakurai2022hybrid}&	Macro hybrid split\\
Hybrid quantum-classical algorithm for hydrodynamics	\cite{zylberman2022hybrid}&	Macro hybrid split\\
Hybrid quantum-classical search algorithms	\cite{rosmanis2022hybrid}&	Macro hybrid split\\
A quantum algorithm of k-means toward practical use	\cite{ohno2022quantum}&	Macro hybrid split\\
A hybrid quantum-classical algorithm for robust fitting. \cite{ doan2022hybrid}&	Macro hybrid split\\
Graph-$|Q >< C|$, a graph-based quantum/classical algorithm for efficient electronic structure on hybrid &\\
quantum/classical hardware systems: Improved quantum circuit depth performance	\cite{zhang2022graph}&	Parallel hybrid\\
Optimization of robot trajectory planning with nature-inspired and hybrid quantum algorithms	\cite{schuetz2022optimization}&	Parallel hybrid\\
A quantum approach for tactical capacity management of distributed electricity generation	\cite{ phillipson2022quantum}&	Parallel hybrid\\
Hybrid quantum-classical unit commitment	\cite{mahroo2022hybrid}&	Breakdown hybrid\\

    \end{tabular}
\end{table*}

In Table ~\ref{tab:class} this classification is shown. In this table, but also in reality, the majority of papers are within the micro and macro hybrid split classes. We could not find any papers within the processing hybrid class that use the terminology hybrid quantum-classical. Papers in this class mostly use the term quantum algorithm. 

\section{Conclusions}
It is expected that quantum computing will always need some form of classical computing to enable the calculations and the execution on the hardware platforms. This is often named hybrid quantum-classical computing. The term hybrid is however diffuse and multi-interpretable. We showed in this paper that in literature this term covers many concepts. Based on this literature and concepts from workflow approach and classical computer science, we distinguished between horizontal and vertical hybrid quantum computing and defined and named various specific types within these classes. This can help researchers and practitioners in quantum computing to make clear what they mean when using the general term `hybrid quantum-classical computing' and can help in developing more concise tools within the quantum computing stack. We do not assume to be complete in our overview and categorisation. We encourage scientists and practitioners to complement this framework as part of future research on this topic. 

\bibliographystyle{IEEEtran}
\bibliography{IEEEabrv,literature}

\end{document}